\documentclass[preprint,12pt,authoryear]{article}

 \usepackage{arxiv}

\usepackage[utf8]{inputenc} 
\usepackage[T1]{fontenc}    
\usepackage{hyperref}       
\usepackage{url}            
\usepackage{booktabs}       
\usepackage{amsfonts}       
\usepackage{nicefrac}       
\usepackage{microtype}      
\usepackage{lipsum}
\usepackage{bm}
\usepackage{graphicx}
\usepackage{multirow}
\usepackage{color,soul}
\usepackage{ulem,xpatch}
\graphicspath{ {./images/} }
\usepackage{float}
\usepackage{setspace}   
\usepackage{xcolor}
\usepackage{amsmath}
\usepackage{amssymb}
\usepackage{blindtext}
\usepackage{lineno}

\usepackage[round]{natbib}   

\definecolor{darkbrown}{rgb}{0.47, 0.27, 0.23}

\title{Coarse-graining of CFD-DEM for simulation of sand production in the modified cohesive contact model}

\author{
 Daniyar Kazidenov \\
  Nazarbayev University \\
  Astana, Kazakhstan\\
  \texttt{daniyar.kazidenov@nu.edu.kz} \\
   \And
 Furkhat Khamitov\\
  Nazarbayev University \\
  Astana, Kazakhstan\\
  \texttt{furkhat.khamitov@nu.edu.kz} \\
  \And
   Yerlan Amanbek \\
   Nazarbayev University \\
   Astana, Kazakhstan\\
  \texttt{yerlan.amanbek@nu.edu.kz} \\
}

\begin{document}
\maketitle
\begin{abstract}
Sand production is an important issue for many hydrocarbon recovery applications in unconsolidated reservoirs. The model using the Computational Fluid Dynamics coupled with Discrete Element Method (CFD-DEM) can capture micro-scale features of sand transport problems. In this study, a coarse-graining approach of 3D CFD-DEM model is developed for the sand production phenomenon using the sample based on the particle size distribution (PSD) from the Kazakhstan reservoir field. The derivation of scaling from a fine to a coarse model is presented rigorously. The original (fine scale) model is validated to the laboratory results including the sand production rate. The results of the original model is compared to the SSW and SSP coarse-gained models. The SSW model results agrees with the result of the sand production rate for the original system. We also observe a good match of the fluid velocity streamline and the produced particle distribution between the original and the coarse models. The speedup of the coarse model is up to 9.4 in the parallelized coarse-grained model.
\end{abstract}

\section{Introduction}

Sand production is one of the most important concerns in the weakly consolidated reservoirs of the  oil and gas production.
It can damage the equipment due to erosion, blocks the flow lines, and prevent the access to the wellbore, thereby reducing oil and gas production. There are three types of sand production: transient, continuous and catastrophic sand production. In the transient sand production, the produced sand mass decreases with time under a constant well production condition. The continuous sand production happens when sand is produced in a constant pattern \citep{veeken1991sand}. In the catastrophic sand production, a high rate of sand influx causes  sudden clogging and/or shutdown of the well \citep{khamitov2022numerical}.\\
Sand production is a coupled process of fluid-solid phase interaction, which mainly involves two mechanisms: mechanical instability, which results in local plastic behavior and destruction of the rock around the cavity, and subsequent transport of sand particles due to the interaction of fluid and particles. In the numerical analysis, coupling of the computational fluid dynamics \citep{versteeg2007introduction} and discrete element method \citep{cundall1979discrete} (CFD-DEM), has become the most preferred instrument to study the sand production phenomenon. \cite{zhu2007discrete} describe that the behavior can be modeled in a micro-mechanical scale using CFD–DEM. \cite{oconnor1997discrete} presented the application of the DEM approach to characterize the the mechanics of sand production. In \cite{sun2018discrete, sun2020grain}, the depletion-induced compaction effects in unconsolidated sands are investigated by introducing a grain-crushing model based on DEM coupled with particle bonded model (BPM).  \\
The Johnson-Kendall-Roberts (JKR) theory, which was initially proposed by \cite{johnson1971surface}, is one of  the commonly used models to describe the cohesive particles in DEM simulations. In the JKR model, the contact surfaces are deformed due to the adhesive force acting in the contact area. As a result, the contact area in the JKR model becomes larger than that described in the Hertz model. Therefore, JKR model is widely used in the applications with soft materials such as screening process \citep{zhou2022calibration}, processing of cohesive bulk materials \citep{roessler2019parameter}, biomanipulation \citep{korayem2014modeling} and powder mixing \citep{madyarov2021understanding}. Recently, \cite{rakhimzhanova2019numerical} have introduced a modified version of the JKR theory to investigate the cementation behavior of the sandstone sample from the Kazakhstan reservoir field. In their work, the material parameters that are described by the modified JKR model were successfully calibrated to the experimental data \citep{shabdirova2016sample}. \cite{rakhimzhanova2022numerical} applied the modified JKR theory to more precisely model the cementation of the sandstone sample in the sand production simulations. \cite{khamitov2021coupled} enriched the previous work by performing 3D CFD-DEM simulation with a large number of particles to investigate the bond breakage and sand production phenomenon in weak cemented sandstone.\\
The CFD-DEM is a combination of the Eulerian-Lagrangian approach initially proposed by \cite{tsuji1993discrete}, which is utilized for modeling of fluid-particle systems. The CFD models the fluid phase by solving the local averaged Navier–Stokes equations. The solid phase is computed by the DEM, which applies Newton's law to track the position and motion of individual particles and the force-displacement law to determine the contact forces from displacements. Although the DEM guarantees accurate modeling, it becomes computationally expensive due to the execution of simulations at a micro-mechanical level. The computational cost of the DEM simulations rises when the number of particles increases in the system. To reduce the computational cost of the DEM simulations various coarse-graining methods have been developed \citep{di2021coarse}. The main idea of the coarse-graining method is to represent the original particles by decreasing the number of particles in the multiscale setting. This could be obtained by scaling the material parameters and size of the fine particles in order to ensure similar mechanical behavior in both systems. At the beginning of elaboration, the following coarse-graining methods were proposed including the imaginary sphere model \citep{sakano2000numerical}, similar particle assembly model \citep{kuwagi2004similar}, and the similarity model \citep{washino2007similarity}. These models were developed by considering only the drag force, in which the particle-fluid interaction force is more significant than other forces of the system. Over the last few years, the coarse-graining approach has been used in a wide range of applications, such as fluidized beds \citep{sakai2010large, mu2020scaling}, a dense medium cyclone \citep{chu2016applicability}, and a pneumatic conveying \citep{sakai2009large}. The latter developed the coarse-grained model by focusing on the contact and drag forces in the particle-fluid and particle-particle interactions. \\
The coarse-graining techniques for the JKR contact model were commonly developed by considering cohesive force and surface energy density. \cite{bierwisch2009three} proposed a quadratic force scaling based on energy density and stresses to simulate the cavity filling with powders. It was concluded that while the scaling based on the energy density is appropriate for dilute systems, the scaling based on the stress based has found better use for dense systems. \cite{chen2020scaling} implemented a scaling method for the JKR model based on mass, momentum and energy conservation. In their proposed approach, the cohesive force and work of adhesive peeling are scaled cubically in the coarse-grained system. \cite{sakai2012study} developed a coarse-grained model to model the cohesive particles by quadratic scaling of the Van der Waals force. In their model, although the inter-surface distance of the particles depends on the coarse-grain factor, the Hamaker constant is assumed to be constant in both systems. \\
This work presents the coarse-grained method of the coupled CFD-DEM model in 3D for the sand production phenomenon. The simulations are performed using the modified cohesive contact model in DEM with the sample based on particle size distribution from the Kazakhstan field. We use the Navier-Stokes flow model as CFD which is coupled to DEM that is solid particle system with the application of  Newton's second law. We derive the scaling rules from the original system (fine-scale system) to the coarse system rigorously taking into account materials properties and the particle distribution. 
The original system with fine particles was validated by the laboratory experimental data. This result of the original system is compared with the results of the proposed coarse-grained models with 8 times less number of particles. In particularly, we focus on the cumulative sand production rate and PSD of the produced particles of the original and coarse-grained models. The computational speedup of coarse-grained models with parallelization setting is reported.

The outline of the paper is organized as follows: in Section 2, we present the mathematical derivations of the coarse-grained approach for the CFD-DEM model. The numerical simulation setting, verification of the original model with the experimental data, and comparison of the coarse-grained models are presented in Section 3. Finally, in Section 4, we summarize the main conclusions of this study.





\section{Numerical model formulation}

In this section, we first describe the particulate model and its coupling with fluid, next introduce the cohesive contact model and finally develop the coarse-graining methods.  

\subsection{Model of the solid phase}

The Discrete Element Method (DEM) is a numerical model that characterizes the mechanical behavior of solid particles, which was initially proposed by \cite{cundall1979discrete}. In this model, Newton’s second law describes the motion of the particles and the force-displacement law is applied to determine the contact forces from displacements.

The translational and rotational motion of the particles can be described as follows: 

\begin{equation}
m\frac{d \bm{v}}{dt} = \bm{F}_{g} + \sum \bm{F}_{c} + \bm{F}_{d} + \bm{F}_{b}\label{eq:1}
\end{equation}

\begin{equation}
I\frac{d \bm{\omega}}{dt} = \sum \bm{T}_{c}\label{eq:2}
\end{equation}

where $\bm{v}$ and $\bm{\omega}$ are the translational and angular velocities of the particle, $m$ and $I$ are the mass and rotational inertia of the particle, $\bm{F}_{c}$ and $\bm{T}_{c}$ the contact force and torque acting on the particle, $\bm{F}_{g}$ is the gravitational force, $\bm{F}_{d}$ is the damping force and $\bm{F}_{b}$ - buoyancy. 

The contact force $\bm{F}_{c}$ is described by the linear spring-dashpot-slider model \citep{cundall1979discrete}, and consists of normal and  tangential components:

\begin{equation}
\bm{F}_{c}^{(n)} = - k_{n} \bm{\delta}_n + \eta_n \bm{v}_n \label{eq:3}
\end{equation}

\begin{equation}
\bm{F}_{c}^{(t)} = - min (\mu \bm{F}_c^{(n)}, k_{t} \bm{\delta}_t + \eta_t \bm{v}_t) \label{eq:4}
\end{equation}

where (sub $n$ - normal, sub $t$ - tangential) $\bm{\delta}$ is the overlap between contacting particles, $\bm{v}$ is their relative velocity at the contact point, $k$ is the spring stiffness constant, $\eta$ is the dashpot damping coefficient, and $\mu$ is the slider friction coefficient. 

The contact torque consists of only tangential torque and it is described as follows:

\begin{equation}
\bm{T}_{c}^{(t)} = \bm{r} \times \bm{F}_{c}^{(t)} \label{eq:5}
\end{equation}
where $r$ is the particle radius.

\subsection{CFD-DEM coupling} 
\cite{tsuji1993discrete} proposed the CFD-DEM technique to simulate the fluid-particle systems. In CFD-DEM, the particulate phase is modeled by DEM, and the fluid phase is solved by CFD, in which locally averaged Navier-Stokes equation calculates the fluid flow in the porous region of the material. In so-called model A \citep{zhu2007discrete}, the fluid occupies only a porous fraction (n) of the material and the pressure drop is divided between fluid and solid phases: 

\begin{equation}
 \left\{
\begin{array}{l}
\rho_f\frac{\partial n \bm{u}}{\partial t} + \rho_f \bm{u}\cdot\nabla(n \bm{u})=  -n\nabla p + \mu_f \nabla^2 (n \bm{u}) +n\rho_f \bm{g} + \bm{f}_d  \\
\vspace{0.01cm} \\
\frac{\partial n}{\partial t} + \nabla(n \bm{u})=  0 \label{eq:6} \\
 \end{array} 
  \right.
\end{equation}

where $\rho$ is the fluid density, $u$ is the fluid dynamic velocity, $p$ is the fluid pressure, $\mu_f$ is the fluid viscosity, and $f_d$ is a particle–fluid interaction force.

The drag force acting on the fluid in each cell can be determined from the sum of drag forces acting on all particles contained in the fluid cell:
\begin{equation}
\bm{f}_{d} = \frac{\sum_{j} \bm{f}_{d}^j}{V_{cell}} \label{eq:7}
\end{equation}

where $\bm{f}_d^j$ is a particle drag force, $V_{cell}$ is the volume of the fluid cell.

The particle drag force is described by the following empirical expression \citep{wen1966mechanics}: 
\begin{equation}
\bm{f}_{d}^j = \left(\displaystyle\frac{1}{8} C_d \rho_f \pi d^2_p |\bm{u}-\bm{v}| (\bm{u} -\bm{v}) \right)n^{-\chi} \label{eq:8}
\end{equation}

where $C_d$ is the drag coefficient, $d_p$ is the particle diameter, $\bm{u}$ and $\bm{v}$ are the fluid and particle velocities, $n$ is the porosity and  $\chi$ is a correction factor. 

The drag coefficient and porosity correction are described by \cite{di1994voidage} empirical correlations: 
\begin{equation}
C_{d} = \left(\displaystyle 0.63 + \frac{4.8}{\sqrt{Re_{p}}} \right)^{2} \label{eq:9}
\end{equation}

\begin{equation}
\chi = 3.7 - 0.65 \exp \left(\displaystyle - \frac{(1.5 - \log_{10} (Re_{p}))^{2}}{2} \right) \label{eq:10}
\end{equation}

where $Re_p$ is a particle Reynolds number that is given by:

\begin{equation}
Re_{p} = \frac{\rho_f d_p|\bm{u}-\bm{v}|}{\mu_f} \label{eq:11}
\end{equation}

\subsection{Description of the Modified JKR Models}

\begin{figure}[h!] 
    \centering
    \includegraphics[width=10cm, height=5.88cm]{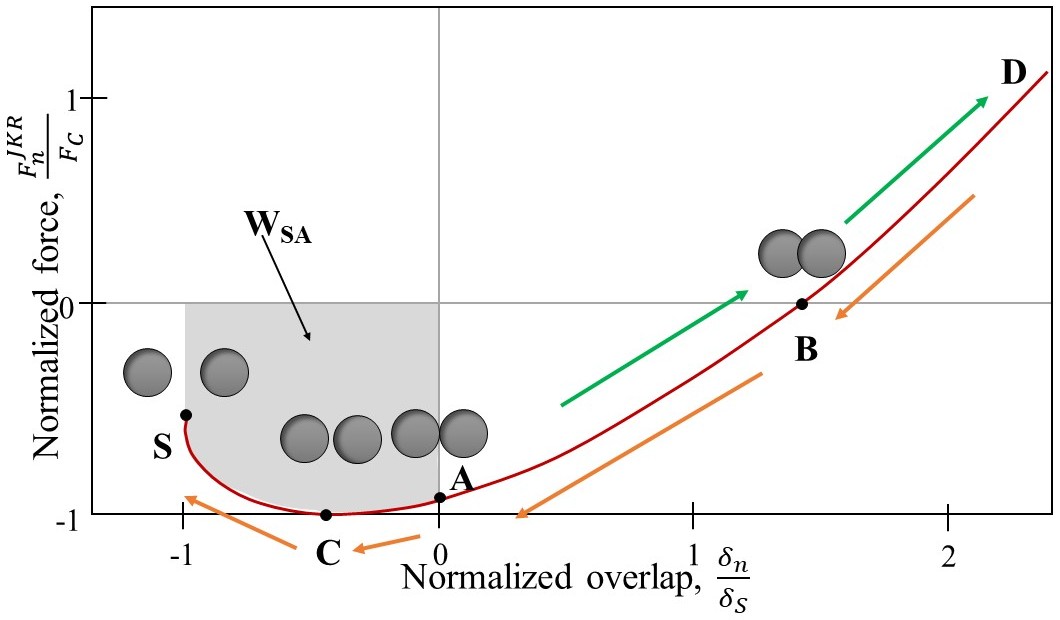}
    \caption{Force-displacement relationship of the JKR model.}
    \label{fig:fig1}
\end{figure}

At the JKR model, two colliding particles establish contact at point \textbf{A} and loading begins, see Figure \ref{fig:fig1}. At this point, the normal force instantly drops to $F_{A} = -\frac{8}{9}F_{C}$, due to the van der Waals attractive forces. The velocity of the particles is decreased moderately, and the overlap (deformation) between the particles is increased. At the point \textbf{B}, the overlap is raised resulting in an increase of the repulsive force between the particles. Therefore, the contact force reaches equilibrium.  At the point \textbf{D}, the overlap reaches its maximum value, completing the compression. Then the unloading process begins. After, the deformation returns to the point \textbf{A} where the overlap returns to a zero value. At the point \textbf{C}, the particles continue to distance each other, where the maximum attractive (pull-off) force reaches its maximum value. Then, with a further decrease in normal force, the contact breaks at the point \textbf{S}, which leads to separation of the particles.

\begin{figure}[h!] 
    \centering
    \includegraphics[width=10cm, height=5.88cm]{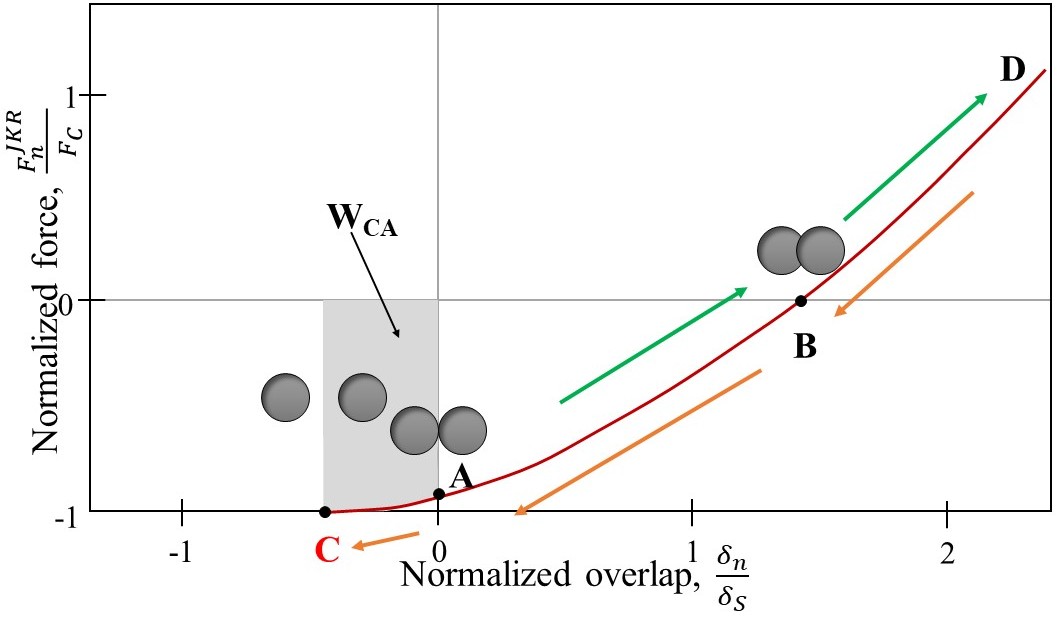}
    \caption{Force-displacement relationship of the modified JKR model.}
    \label{fig:fig2}
\end{figure}

\cite{rakhimzhanova2019numerical} proposed a modified version of the JKR model for weak cemented sandstone in Kazakhstan, see Figure \ref{fig:fig2}. Two assumptions were implemented in their model: bond breakage at the maximum value of the normal contact force and no new bonds formation after breakage. This is characterized by the fact that the bonds in cemented sandstone are not tensile and break in a brittle fashion. It is assumed that bonds break at point \textbf{C} instead of point \textbf{S}. New contacts become unbonded and are described by the Hertz model.

According to \cite{johnson1971surface}, the contact force in the normal direction between the cohesive particles is calculated as follows:

\begin{equation}
F_{n} = \frac{4E^{*}a^{3}}{3R^{*}}-\sqrt{16\pi\gamma E^{*}a^{3}}\label{eq:12}
\end{equation}
where, $E^{*}$ is the effective Young's modulus given by $E^{*} = \left(\displaystyle \frac{1 - \nu^{2}_{1}}{E_{1}} +\displaystyle \frac{1 - \nu^{2}_{2}}{E_{2}}\right)^{-1}$, where $E_{1}$, $E_{2}$ and $\nu_{1}$, $\nu_{2}$ are Young's modulus and Poisson's ratios of the colliding particles. $R^{*}$ is the effective radius expressed by $R^{*} = \left(\displaystyle \frac{R_{1}R_{2}}{R_{1}+R_{2}}\right)$, where  $R_{1}$ and $R_{2}$ are the radii of the particles, $\gamma$ is $a$ surface energy density and  a is a contact radius. 

The overlap between particles can be computed as follows:

\begin{equation}
\delta = \displaystyle \frac{a^{2}}{R^{*}}-\sqrt{\displaystyle \frac{4\pi\gamma a}{E^{*}}}\label{eq:13}
\end{equation}

The forces, overlaps, and contact radii of the particles at above-mentioned points are presented in Table \ref{table:1}.

\begin{table}[h]
  \caption{Force, overlap, contact radius at points C, A, B \citep{chen2020scaling}}
  \centering
  \begin{tabular}{l c c c}
    \toprule
    Point     & Force     & Overlap  & Contact radius  \\
    \midrule
    C     &  $ - 3\pi \gamma R^{*}$  & $-0.57 \left(\displaystyle \frac{\pi^{2} \gamma^{2} R^{*}}{E^{*2}} \right)^{\displaystyle \frac{1}{3}}$ &  $\left(\displaystyle \frac{9\pi \gamma R^{*2}}{4E^{*}} \right)^{\displaystyle \frac{1}{3}}$    \\
    A     & $- 2.67\pi \gamma R^{*}$    & $0$ & $\left(\displaystyle \frac{4\pi \gamma R^{*2}}{E^{*}} \right)^{\displaystyle \frac{1}{3}}$  \\
    B     & $0$    & $1.44 \left(\displaystyle \frac{\pi^{2} \gamma^{2} R^{*}}{E^{*2}} \right)^{\displaystyle \frac{1}{3}}$  & $\left(\displaystyle \frac{9\pi \gamma R^{*2}}{E^{*}} \right)^{\displaystyle \frac{1}{3}}$  \\
    \bottomrule
  \end{tabular}
  \label{table:1}
\end{table}

\subsection{Coarse-Graining Approach}

In the coarse-graining method, the original particles in the simulation are replaced with their representatives, which are in less number. They are defined as grains or coarse-grained particles. The simulation time of the modeling is saved due to the particle size enlargement. Therefore, this leads to a decrease in the total number of particles in the system. However, the behavior of these particles, including the conservation of mass, the momentum, and the energy conservation must be the same as those of the original ones. The coarse-graining technique is described by a coarse-grain factor and coarse-grain number. The coarse-grain factor shows the ratio between the diameters of the coarse-grained particle to the original particle:
\begin{equation}
k = \frac{d_{g}}{d_{p}}\label{eq:14}
\end{equation}
where $d_{g}$ and $d_{p}$ are the diameters of the coarse-grained particle and original particle. Subscripts $g$ and $p$ define variables related to coarse-grained and original particles, respectively. The coarse-grain number $n_{CG}$ describes the number of original particles per grain:
\begin{equation}
n_{CG} = \frac{n_{p}}{n_{g}}\label{eq:15}
\end{equation}
\begin{equation}
n_{CG} = k^{3}\label{eq:16}
\end{equation}
where $n_{p}$ and $n_{g}$ are numbers of the particles and grains. The volume of the coarse-grained system and the density of the grains must be the same as those of the original ones in order to ensure that the mass of both systems is conserved. This can be seen from the following:
\begin{equation}
M_{tot} = \sum_{n_{g}} m_{g} = \sum_{n_{p}} m_{p}\label{eq:17}
\end{equation}
\begin{equation}
V_{tot} = \sum_{n_{g}} V_{g} = \sum_{n_{p}} V_{p}\label{eq:18}
\end{equation}
\begin{equation}
\rho_{g} = \rho_{p}\label{eq:19}
\end{equation}
\begin{equation}
m_{g} = k^{3}m_{p}\label{eq:20}
\end{equation}
where $M_{tot}$ and $V_{tot}$ - system mass and volume, $m_g$, $m_p$  $V_g$, $V_p$, $\rho_g$ and $\rho_p$ - mass, volume and density of the single grain and particle, respectively.

In the coarse-graining method, particle size scaling alone is insufficient to provide a similar mechanical behavior for the original particles and grains. Therefore, the material parameters of the original particles  must also be scaled to represent them in the coarse-grained system. We focus on the scaling  in respect of material characteristics and dispersity of the particles in the system.  

In terms of material characteristics, we examined an appropriate scaling method for the dense medium system. This is due to the fact that the system in the current simulation consists of compressed particles. For such a system, \cite{chu2016applicability}  proposed quadratic scaling for the interparticle force by the conservation of momentum.
In order to represent the original system with coarse-grained system, the total energy, which includes potential and kinetic energy, should be the same in both systems. Potential energy is conserved since mass of the systems is equal. To conserve kinetic energy the velocity of the particles and grains should be the same because mass is conserved in both systems.
According to the impulse-momentum change equation ($m\bm{v} = \bm{F}t$) and equation (\ref{eq:20}), the acting forces in both systems should be the same: 

\begin{equation}
\bm{F}_{g}t_{g} = k^{3} \bm{F}_{p}t_{p}\label{eq:21}
\end{equation}
where $\bm{F}_g$ and $\bm{F}_p$ are the total acting forces in coarse-grained and original system, respectively.

Since the diameters of the particles scales as $d_{g} = kd_{p}$ at the same velocities, the acting time of the forces between original particles and grains scales as $t_{g} \approx kt_{p}$.  Therefore, the contact force for particle-particle interaction can be computed as:
\begin{equation}
F_{c,g} = k^{2}F_{c,p}\label{eq:22}
\end{equation}
This result was also obtained in \cite{bierwisch2009three}.
From equation (\ref{eq:22}) and a break force $F_{C}$ given in Table \ref{table:1}, the scaling law for the surface energy density can be computed as follows:
\begin{equation}
F_{C,g} = k^{2}F_{C,p}\label{eq:23}
\end{equation}
\begin{equation}
\gamma_{g} = k\gamma_{p}\label{eq:24}
\end{equation}

The work performed during the breakage of two adhered particles is defined as  work of adhesive peeling. It can be indicated by the shaded area between points \textbf{C} and \textbf{A} in Figure \ref{fig:fig2}.  The work of adhesive peeling $W_{CA}$ is calculated by integrating the contact force by the overlap in that area: 

\begin{equation}
W_{CA} =  \int_{\delta_{A}}^{\delta_{C}}F_{n}d\delta \label{eq:25}
\end{equation}

By inserting equations (\ref{eq:12}) and (\ref{eq:13}) into equation (\ref{eq:25}) , we obtain:

\begin{equation}
W_{CA} = \int_{a_{A}}^{a_{C}} \left(\displaystyle \frac{4E^{*}a^{3}}{3R^{*}} - \sqrt{16\pi\gamma E^{*}a^{3}} \right)\left(\displaystyle \frac{2a}{R^{*}} - \sqrt{\displaystyle \frac{\pi\gamma}{E^{*}a}} \right)da = 11.12\left(\displaystyle \frac{\gamma^{5}R^{*4}}{E^{*2}} \right)^{\displaystyle \frac{1}{3}} =0.965F_{C}\delta_{C} \label{eq:26}
\end{equation}

Since the work of adhesion is performed in a cohesive contact model, it is expressed as a function of material parameters such as surface energy density, radius and Young's modulus. In addition, it can be described by force and in terms of force and overlap. In this case, from equation (\ref{eq:26}) it can be seen that the work is equal to the product of the breaking force and the overlap at point \textbf{C}. 

\cite{chen2020scaling} proposed scaling rule for the work of adhesive peeling  by applying energy balances of the original particle and coarse-grained particle systems:
\begin{equation}
W_{g} = k^{3}W_{p}\label{eq:27}
\end{equation}

By substituting equation (\ref{eq:26}) into (\ref{eq:27}), it is clear that the effective Young's modulus is the same in both systems:

\begin{equation}
E^{*}_{g} = E^{*}_{p}\label{eq:28}
\end{equation}

Since the effective Young's modulus is the function of Poisson's ratio and Young's modulus, it is evident that both Poisson's ratio and Young's modulus are the same in original and coarse-grained systems when the material of the colliding particles in the system is the same i.e. $\nu = \nu_{1} = \nu_{2}$ and $E = E_{1} = E_{2}$, 

According to Table \ref{table:1} and equations (\ref{eq:24}), (\ref{eq:28}), we can scale the contact overlap and contact radius for the coarse- grained system:
\begin{equation}
\delta_{g} = k\delta_{p}\label{eq:29}
\end{equation}
\begin{equation}
a_{g} = ka_{p}\label{eq:30}
\end{equation}

In order to scale the restitution coefficient $e$, the normal component of the damping force is balanced between two systems. As with the contact force, the damping force is also scaled quadratically: 

\begin{equation}
F_{d,g} = k^{2} F_{d,p}\label{eq:31}
\end{equation}

By substituting the damping forces with their components accordingly, we can state that the restitution coefficient $e$ remains unchanged in both systems: 

\begin{equation}
\displaystyle -\beta_{g} \sqrt{\displaystyle \frac{20}{3} E^{*}_{g}a_{g}m^{*}_{g} u_{n,g}} = \displaystyle -k^{2}\beta_{p} \sqrt{\displaystyle \frac{20}{3} E^{*}_{p}a_{p}m^{*}_{p} u_{n,p}}\label{eq:32}
\end{equation}

From equations (\ref{eq:20}), (\ref{eq:30}) and (\ref{eq:32}), we obtain $\beta_{g} = \beta_{p}$, i.e.

\begin{equation}
\frac{\ln e_{g}}{\sqrt{\ln e_{g} + \pi^{2}}} = \frac{\ln e_{p}}{\sqrt{\ln e_{p} + \pi^{2}}}\label{eq:33}
\end{equation}
Therefore, we have
\begin{equation}
e_{g} = e_{p}\label{eq:34}
\end{equation}
where $\beta$ is the critical damping ratio ($\beta = 0$ when there is no damping, $\beta = 1$ refers to critical damping of the system), $u_{n}$ is the normal velocity of the particle (grain), $m^{*}$ is the effective mass given by $m^{*} = \frac{m_{1}m_{2}}{m_{1}+m_{2}}$ , where $m_{1}$ and $m_{2}$ are the masses of the two interacting particles (grains).

Finally, the friction coefficient $\mu$ can be scaled using the Coulombs sliding limit:

\begin{equation}
F_{t} \leq \mu F_{n}\label{eq:35}
\end{equation}
where $F_{t}$ is the tangential component of the contact force. Since $F_{t}$ is scaled quadratically as the normal component, the friction coefficient is the same for grains and original particles.

We derived scaling laws by considering the material characteristics of the system. Summary of scaling laws is provided in Table \ref{tab:table3}. It is apparent that only the surface energy density changes when the simulation is performed in the coarse-grained model. 

\begin{figure}[H] 
    \centering
    \includegraphics[width=13cm, height=6.3cm]{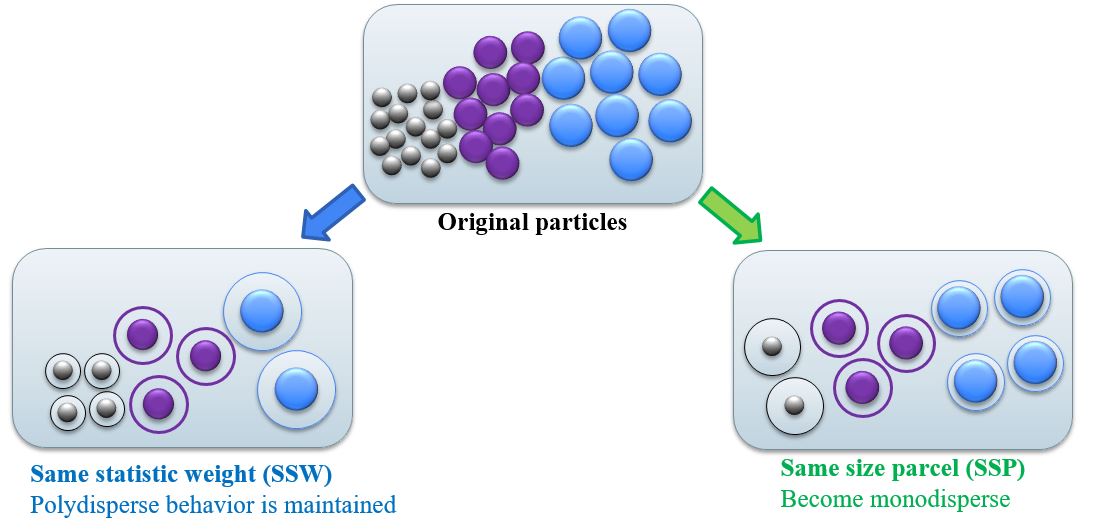}
    \caption{Illustration of the SSW and SSP methods \citep{lu2018assessment}.}
    \label{fig:fig3}
\end{figure}

\cite{lu2018assessment} introduced two different coarse-graining techniques based on dispersity of particles in the system: the same statistic weight (SSW) and the same size parcel method (SSP). The schematic diagram of the methods is shown in Figure \ref{fig:fig3}. In the SSW method, the size of the particles is directly scaled by a single coarse-grain factor. Therefore, the polydisperse behavior of the system is maintained in the coarse-grained model. This model can precisely reproduce the original particle system in terms of dispersity, but with an increased particle size. In the SSP method, the polydisperse particles are formed into a monodisperse particle system by applying different coarse-grain factors to average the size of the particles. Due to the geometrical simplicity of the particles in this method, the computational cost is expected to reduce drastically.
Using above derived material characterization transformation, we conduct numerical scaling analysis with both polydisperse and monodisperse particles in the system.

\section{Numerical results}
\subsection{Numerical setup of the simulation}

\begin{figure}[h!] 
    \centering
    \includegraphics[width=13cm, height=5.95cm]{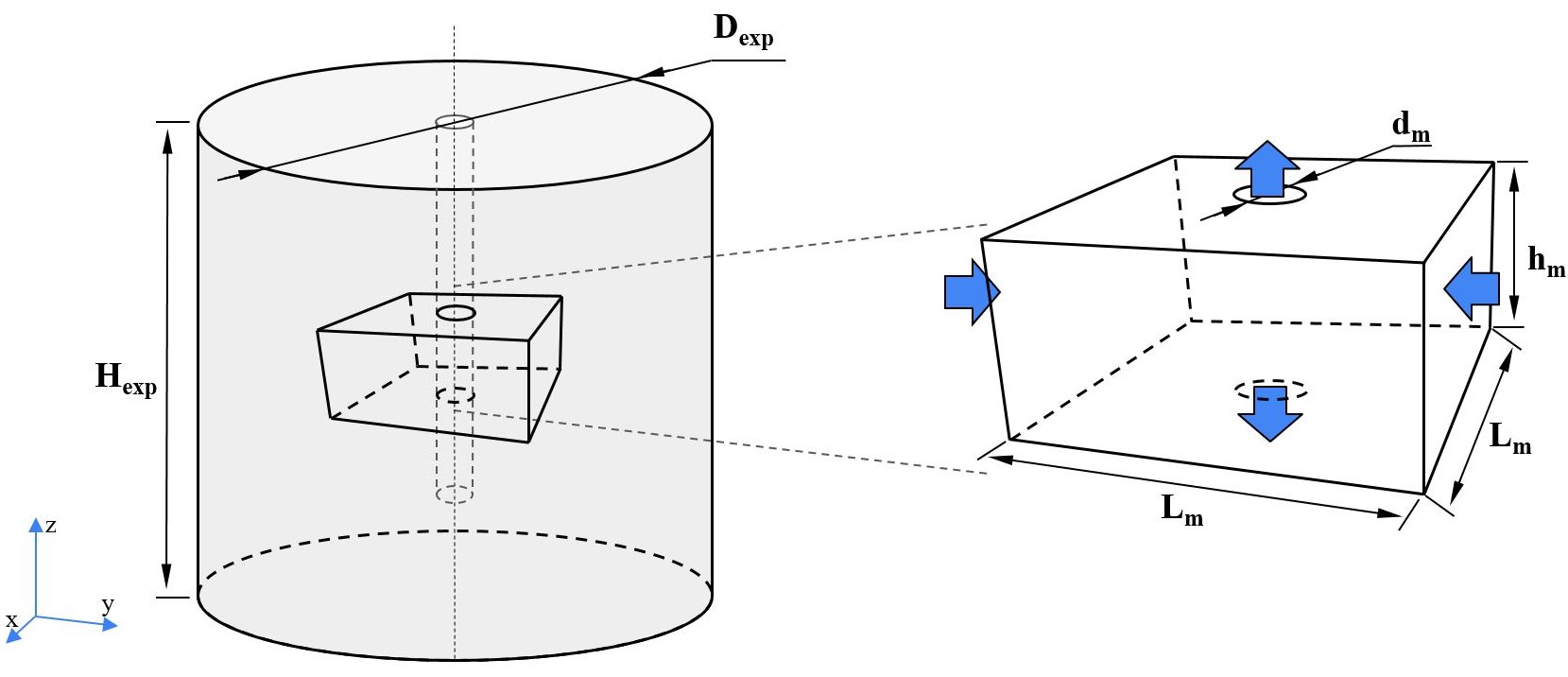}
    \caption{Schematic illustration of the modeling section.}
    \label{fig:fig4}
\end{figure}

In our CFD-DEM model, we have reproduced the real laboratory experiment on the sand production from \cite{kozhagulova2020experimental}. Figure \ref{fig:fig4} shows the schematic illustration of the modeling geometry. The sample size in the experiment is $D_{exp}$ = 300 mm in diameter and $H_{exp}$ = 240 mm in height. It demands the computational power to model the entire laboratory sample, because it contains approximately $10^{9}$ particles. Therefore, we focus on the rectangular box in the center of the sample, where the penetration hole passes, see Figure \ref{fig:fig4}. Then we scaled down it 10 times to have a length $L_m$ = 18 mm and height $h_m$ = 9 mm in the model. The diameter of the particle outlet hole $d_m$, which is located on the top, is equal to 2.6 mm. In this numerical box we could contain 270 000 of original (fine) particles, which is within capability of the computational workstation.

The sand production simulation was performed in Aspherix software \citep{kloss2012models}. All computations were performed on the following High-Performance Computing (HPC) system: Intel(R) Xeon(R) Gold 6230R CPU @ 2.10 GHz with total 104 threads.  In this setting, we considered k = 1 and k = 2 coarse-grain factors, where k = 1 case corresponds to the original system with the polydisperse particle behavior. The coarse-grain factor k = 2 was utilized in the simulations with SSW and SSP methods, in which the first is a polydisperse and second is a monodisperse particle system, respectively. All cases consisted of 4 stages, namely a particle generation, a compression, a perforation and a sand production. Since the first three stages included only solid phases, they were modeled only by DEM. The last stage consisted of two phases: solid and liquid(single-phase fluid), which requires the CFD-DEM coupling system.

\begin{figure}[h] 
    \centering
    \includegraphics[width=10cm, height=7.5cm]{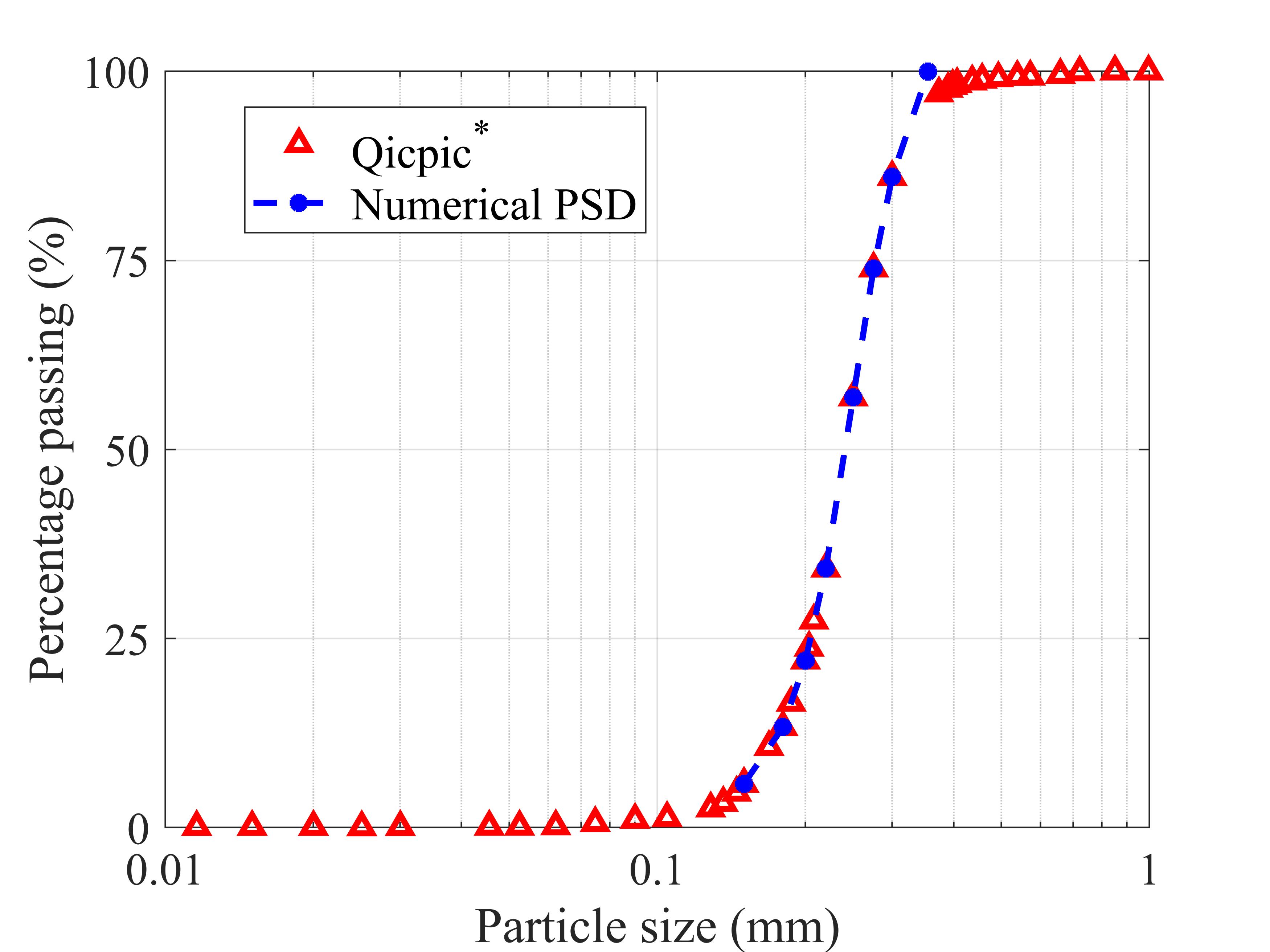}
    \caption{Particle Size Distribution of the real reservoir sand* \citep{shabdirova2016sample} and numerical sample.}
    \label{fig:fig5}
\end{figure}

The geometry of the CFD-DEM coupling are consists of two separate CFD and DEM geometries. Each of them has its own size, mesh cells, meshing type and location in the coordinate system. The size and location in the coordinate of the CFD and DEM geometry domains should be the same during the coupling. 
The DEM geometry domain was constructed as cuboidal cell by 6 square planes with size of 18 mm, in which the particles were first generated. At this stage, the particle sizes for original particle simulation were selected in accordance with the particle size distribution (PSD) of the reservoir sand from the Ustyurt-Buzachi Sedimentary Basin, Kazakhstan \citep{shabdirova2016sample}, see Figure \ref{fig:fig5}. In total, 8 random points are chosen, which are blue dots, to reproduce the real PSD in the modeling. Similar PSD of the numerical sample for CFD-DEM models was used in \cite{shabdirova2020experimental, rakhimzhanova2019numerical,rakhimzhanova2021numerical, khamitov2021coupled}. The PSD by mass fraction of the original and coarse-grained systems are provided in Table \ref{tab:table2}. Since the SSP contains the monodisperse particles, its particle size was calculated by multiplying the coarse-grain factor to the average particle diameter of the original system.

\begin{table}[h!]
 \caption{Particle Size Distribution of the particles used in these simulations}
  \centering
\begin{tabular}{ l l l l } 
\hline
\multirow{2}{4em}{Mass fraction} & \multicolumn{3}{c}{Diameter of the particle (mm)} \\ 
\cline{2-4}
& Original particle  & SSW (k = 2) & SSP (k = 2) \\
\hline
 0.058   & 0.15  & 0.3 &  \multirow{8}{2em}{0.46}\\
 0.075   & 0.18  & 0.36  & \\
 0.088   & 0.2   & 0.4  &  \\
 0.1216  & 0.22  & 0.44  &  \\
 0.2264  & 0.25  & 0.5   & \\
 0.1705  & 0.275 & 0.55  & \\
 0.121  & 0.3   & 0.6   & \\
 0.1395 & 0.355 & 0.71  & \\
 \hline
 \end{tabular}
   \label{tab:table2}
\end{table}

The material parameters of the simulated particles in original and coarse-grained systems can be found in Table \ref{tab:table3}. The parameters of the coarse-grained particles are scaled accordingly using the scaling laws derived above. Similar material parameters were also used in the investigations of \cite{kazidenov2022coarse}, in which the triaxial compression test was performed with cohesive particles in 3D DEM. In their study, they developed a coarse-graining model that could work with polydisperse particles and showed a good agreement in the stress-strain curve with results of the original particle system.     

\begin{table}[h!]
 \caption{Material and scaling parameters of the solid phase used in the simulations}
  \centering
\begin{tabular}{l l l l l} 
\hline
 Coarse-grain factor  & Scaling law & k = 1  & k = 2   \\
\hline
 Density $(kg/m^{3})$         & $\rho_{g}/ \rho_{p} =1$  & \multicolumn{2}{c}{2605}   \\
 Young's modulus $(Pa)$       & $E_{g}/ E_{p} =1$ &   \multicolumn{2}{c}{$2\cdot10^{10}$}      \\ 
 Poisson's ratio              & $\nu_{g}/ \nu_{p} =1$&  \multicolumn{2}{c}{0.3}  \\
 Restitution coefficient      & $e_{g}/ e_{p} =1$ & \multicolumn{2}{c}{0.8} \\
 Friction coefficient         & $\mu_{g}/ \mu_{p} =1$  &\multicolumn{2}{c}{0.2} \\
 Surface energy $(J/m^{2} )$  & $\gamma_{g}/ \gamma_{p} =k$ & 30   & 60   \\
 Particle (grain) number      & $N_{g}/N_{p} = 1/k_{3}$ & 270 000 & 33 750 \\
 \hline
 \end{tabular}
   \label{tab:table3}
\end{table}

The main purpose of the particle generation stage is to produce and prepare the particles for the next further stages. At this stage, we do not consider any bonding effects between particles. Therefore, the original JKR contact model was applied for the particle-particle interaction, and the Hertz contact model was chosen for the particle-wall interaction. 

After the particle generation, the particles were compressed with overburden stress of 1 MPa moving the top plate down. At the final stress the porosity of the sample resulted in 41 \%, which consistent with the laboratory experiment data \citep{kozhagulova2020experimental}. Here the domain length/width remained the same and the height levelled at about 9 mm. From this point, the size of the domain was fixed for further simulation stages.

\begin{figure}[h!] 
    \centering
    \includegraphics[width=16cm, height=9.19cm]{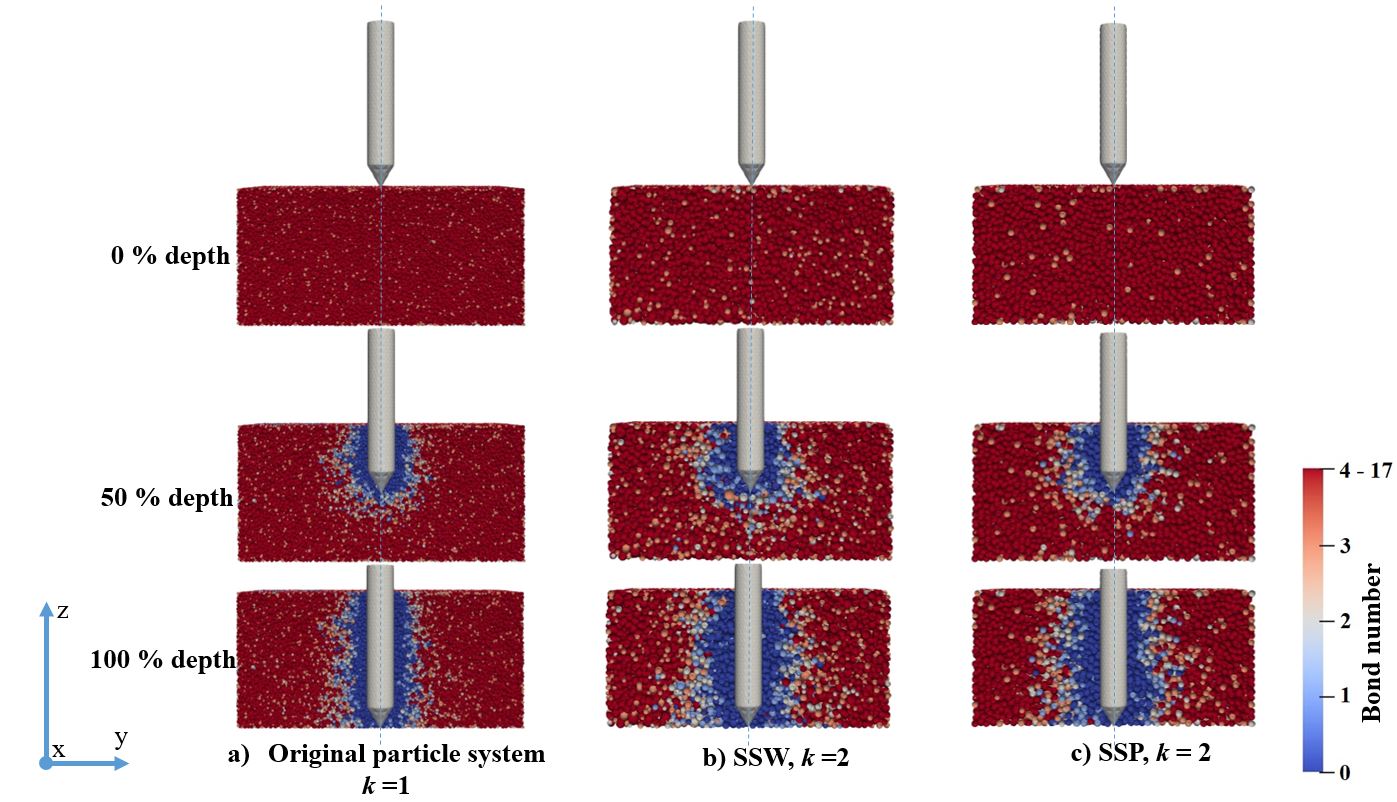}
    \caption{Perforation of the numerical sample in a) Orignial particle system b) SSW and c) SSP models.}
    \label{fig:fig6}
\end{figure}

Next, the numerical sample was perforated with 1.7 mm in diameter and 10 mm in height penetrometer. In this stage, the modified JKR model was applied. Figure \ref{fig:fig6} shows a frontal cross-section view of the numerical sample perforation at 0\%, 50 \% and 100 \% of the sample depth. The color bar on the right determines the number of bonds per particle. A blue color indicates the absence of the bonds between particles, i.e the bonds that are broken. Bonds from 4 to 17 are colored red in order to show evident places where the particles are strongly bonded. The perforation was performed vertically from the top center of the sample to the bottom side. The moving force of the penetrometer was constant and adjusted in such a way to ensure gradual perforation without rapid destruction of the sample. Then, as the penetrometer touched the bottom side, it was removed by pulling it up in the reverse direction with the same velocity. 

After the removal of the penetrometer, the CFD-DEM coupling was simulated to initiate the sand production by a fluid injection. In these simulations, water was selected as a fluid, and its properties are presented in Table \ref{tab:table4}.  
\begin{table}[h!]
 \caption{Physical properties of the fluid}
  \centering
\begin{tabular}{l l} 
\hline
 Fluid type  &  Water \\
\hline
 Density $(kg/m^{3})$         &   998.2 \\
 Dynamic viscosity $(Pa\cdot s)$     & 0.001     \\         
 Reference (absolute) pressure  $(Pa)$ & 101 325 \\ 
 \hline
 \end{tabular}
   \label{tab:table4}
\end{table}

\begin{figure}[h!] 
    \centering
    \includegraphics[width=15cm, height=5.65cm]{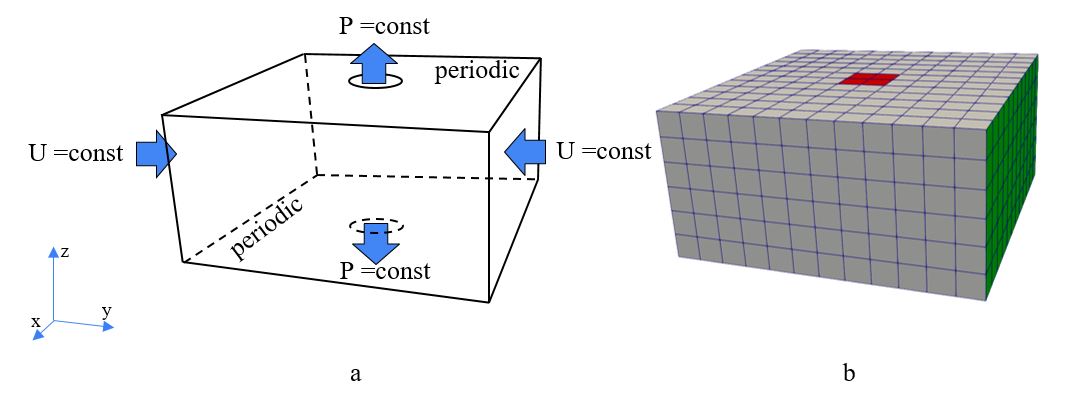}
    \caption{a) Boundary conditions and b) meshing of the CFD part. The red zone is P = const, the green zone is U = const, and the grey zone is periodic boundaries.}
    \label{fig:fig7}
\end{figure}

 Assuming that particles are directed to the upper side due to the effect of fluid flow, the DEM model consisted of a single outlet which was located on the top. In CFD, we defined top and bottom holes as the outlets for the fluid. The size of the CFD domain was the same with the DEM geometry which was used in penetration stage, see Figure \ref{fig:fig7}.  The fluid with velocity of $U = 10^{-4} m/s$ was injected from the plates (right and left) which are normal to $y$-axis, faces that are colored to green. Two plates normal to $x$-axis, top and bottom plates were set to periodic boundary conditions, a grey color in the figure. The pressure of the top and bottom holes was equal to absolute value, which is an atmospheric pressure $P_{atm} = 101 325$ Pa, indicated in a red color, as can be seen in the figure. The three dimensional mesh of the CFD part was divided into 12 cells in $x$ and $y$ directions, and 6 cells in $z$ direction. The cell size was equal to 1.5x1.5x1.5 mm. The meshing of the CFD part was performed in such a way, that we could use an unresolved simulation, in which the particle size is much smaller than the cell size of the fluid phase. Therefore, one fluid cell could contain several particles in it. In our case, one cell accommodated 6.52 particles in original particle simulation and 3.26 particles in the coarse-grained simulation, which satisfied the requirement for the unresolved case \citep{clarke2018investigation}: 
  \begin{equation}
 \frac{x_{cfd}}{\bar{d_p}} > 3 \label{eq:36}
 \end{equation}
 where $x_{cfd}$ is the length of the single fluid cell and $\bar{d_p}$ is the average diameter of the particle. 
 
 In the CFD-DEM coupling, the selection of the simulation time-step is essential, because it directly affects the final results. As the geometry, the time-step of the CFD-DEM coupling is defined and calculated separately. The DEM time step is described by Rayleigh critical time-step and computed using the following equation from \cite{li2005comparison}: 
  \begin{equation}
\Delta t_{c} = \frac{\pi \bar{R}}{\beta} \sqrt{\frac{\rho_p}{G}} \label{eq:37}
\end{equation}

where $\bar{R}$ is the average particle radius, $\rho_p$ is the density of the particles, $G$ is the particle shear modulus, which is $G = \frac{E}{2(1-\nu)}$ and $\beta$ can be computed by:
 \begin{equation}
\beta = 0.8766 + 0.163 \nu \label{eq:38}
\end{equation}

The DEM time step in the current simulations was $10^{-8}$ s, and it was 4.4 \% of Rayleigh critical time-step according to equation (\ref{eq:35}). The CFD time step was equal to $10^{-6}$ s, resulting in a maximum Courant number of 0.0001.

\subsection{Verification of the model}

First, the sand production results of the original particle model were verified with the experimental data \citep{kozhagulova2020experimental} and compared with other studies such  the CFDEM model \citep{khamitov2021coupled} and semi-analytical model \citep{shabdirova2020role}. 
We compared our result with the previously obtained results for various models that use the dimensionless parameters.
The dimensionless time period is computed as follows:
\begin{equation}
t_{d} = \left(\displaystyle \frac{t}{t_{end}}\right)\label{eq:39}
\end{equation}
where $t$ - current time, $t_{end}$ - final time, where the sand production ends. The dimensionless cumulative sand production is defined as:
\begin{equation}
M^{t}_{d} = \left(\displaystyle \frac{ \int_{t=0}^{t}M_{t}dt}{\int_{t=0}^{t_{end}}M_{t}dt}\right)\label{eq:40}
\end{equation}
where $M_{t}$ - sand production mass in kg. 

\begin{figure}[h!] 
    \centering
    \includegraphics[width=10cm, height=7.39cm]{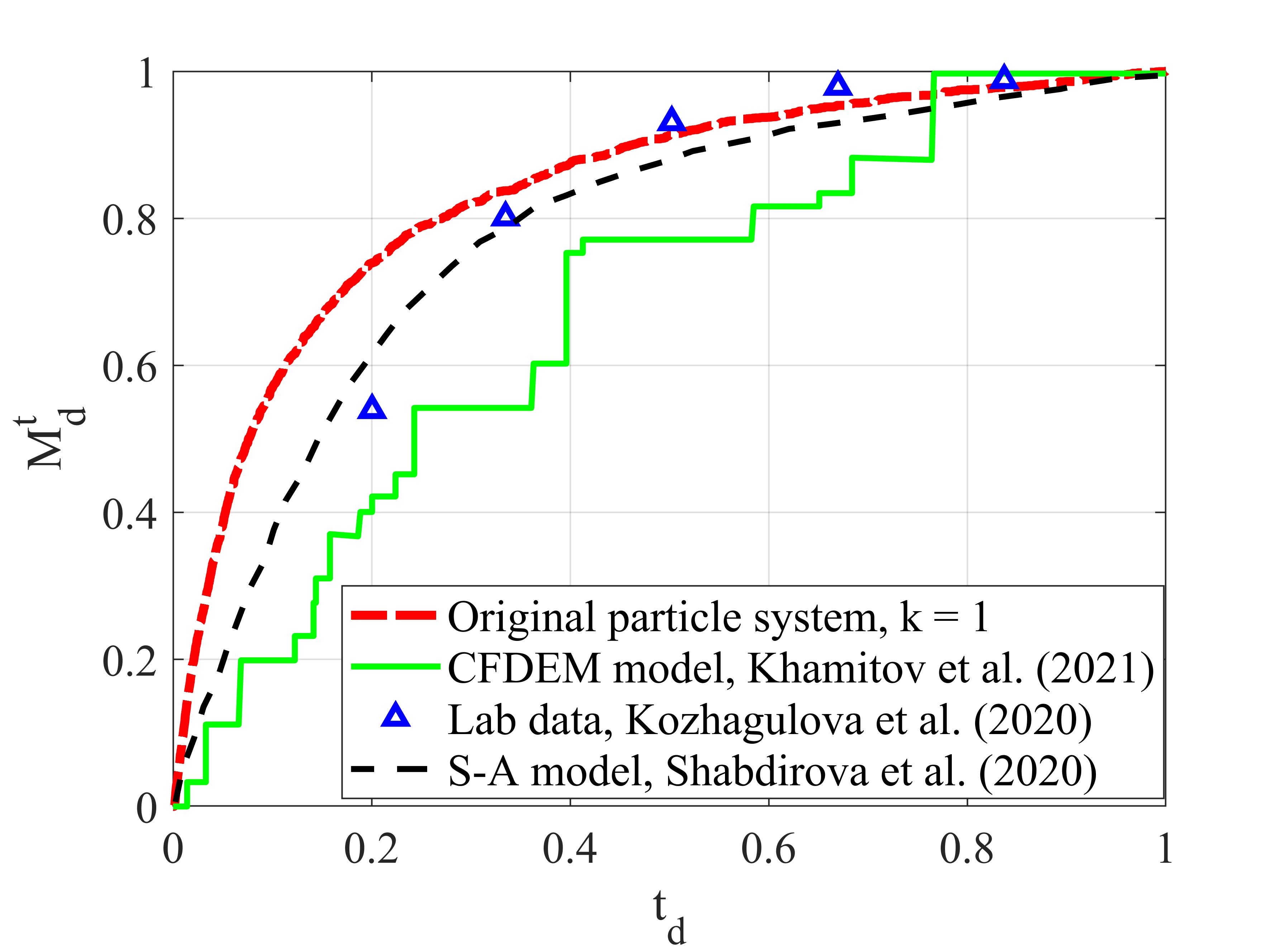}
    \caption{Comparison of the original particle simulation results with the results from different studies.}
    \label{fig:fig8}
\end{figure}

Figure \ref{fig:fig8} demonstrates the verification of the original particle simulation results with the laboratory experimental data. In addition, the CFDEM and semi-analytical models are included for comparison. It can be seen that the original particle model is in good agreement with the experimental result, even the latter is limited to only 5 data.  As can be seen from Figure \ref{fig:fig8}, the original particle and semi-analytical curves are very similar to the laboratory data, and fitted better than previous CFDEM results. Overall, the simulation results of the original particle model show a similar pattern with the experiment in normalized sand production rate. 

In addition, we calculated the root mean squared relative error (RMSRE) to examine the difference between obtained results. The RMSRE is given as follows:
\begin{equation}
RMSRE = \sqrt{\frac{1}{n}\sum_{i=1}^n \frac{M_{d,{exp}}^i - M_{d,{model}}^i}{M_{d,{exp}}^i}}\times 100\label{eq:41}
\end{equation}
where $M_{d,{exp}}^i$ - the $i$-th dimensionless cumulative sand production of the experiment and $M_{d,{model}}^i$ - the $i$-th dimensionless cumulative sand production of the model.

According to equation (\ref{eq:41}), we have obtained the following RMSRE values: for the original - 15.3 \%, semi-analytical model - 6.98 \% and CFDEM model - 19.28 \%.  

\subsection{Comparison of the cumulative sand production between the original and coarse-grained particle models }
\begin{figure}[h!] 
    \centering
    \includegraphics[width=16cm, height=20.658cm]{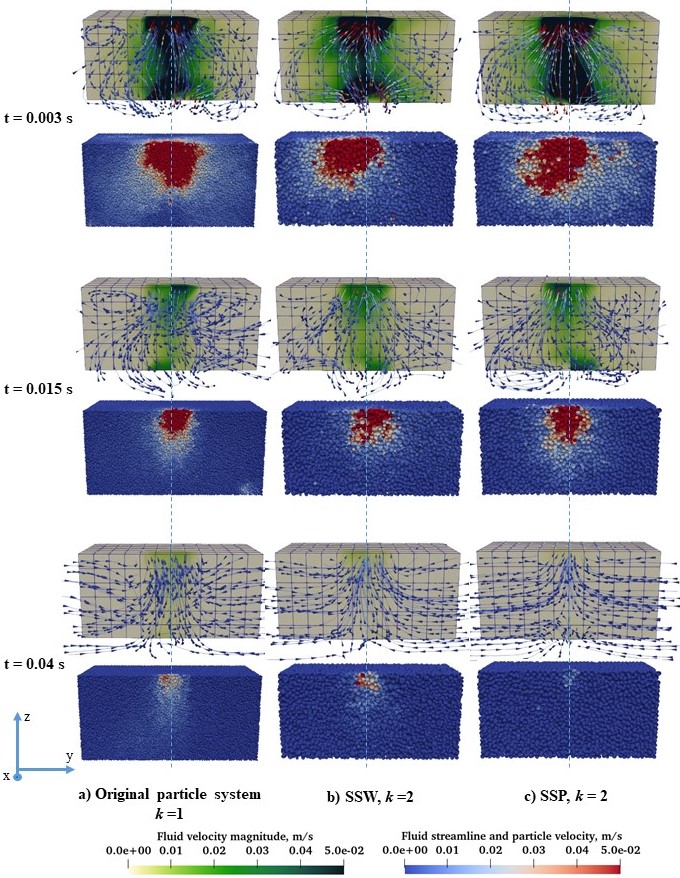}
    \caption{Fluid streamlines, fluid velocity magnitude and particle velocity of a) Original partilce system, b) SSW and c) SSP at t = 0.003 s, t = 0.015 s and t = 0.04 s. }
    \label{fig:fig9}
\end{figure}

Now, we can compare results from coarse-grained models to the result of the verified original model. There was a comparison of the cumulative sand production rate between the original particle and coarse-grained models. The total simulation time was 0.04 s for all cases. During this period the sand production became in transient behavior, that is the produced sand mass decreased over time. Figure \ref{fig:fig9} shows the fluid streamlines with velocity magnitude background and particle velocity in half-sectional view of the numerical sample. The snapshots were captured after 0.003 s, 0.015 s, and 0.04 s from the initial state. At the beginning, we observed a higher velocity magnitude of the fluid at the place where the perforation was performed. The streamlines were curved in the middle of their direction from the bottom hole to the top hole. This phenomenon resulted in the intensive movement of the particles toward the top and consequent their production. At t = 0.015 s, the velocity of the fluid substantially decreased compared to the beginning. This led to a slowdown in sand production rate, since there was not a significant movement of particles toward the top outlet. At the final time, the fluid flow became stationary and streamlines uniformly distributed in the sample. As result no sand production occurred at this time. From the fluid velocity magnitude  behavior and streamline trajectories, it is apparent that the fluid motion in the sample is similar to all models.

\begin{figure}[h!] 
    \centering
    \includegraphics[width=10cm, height=7.5cm]{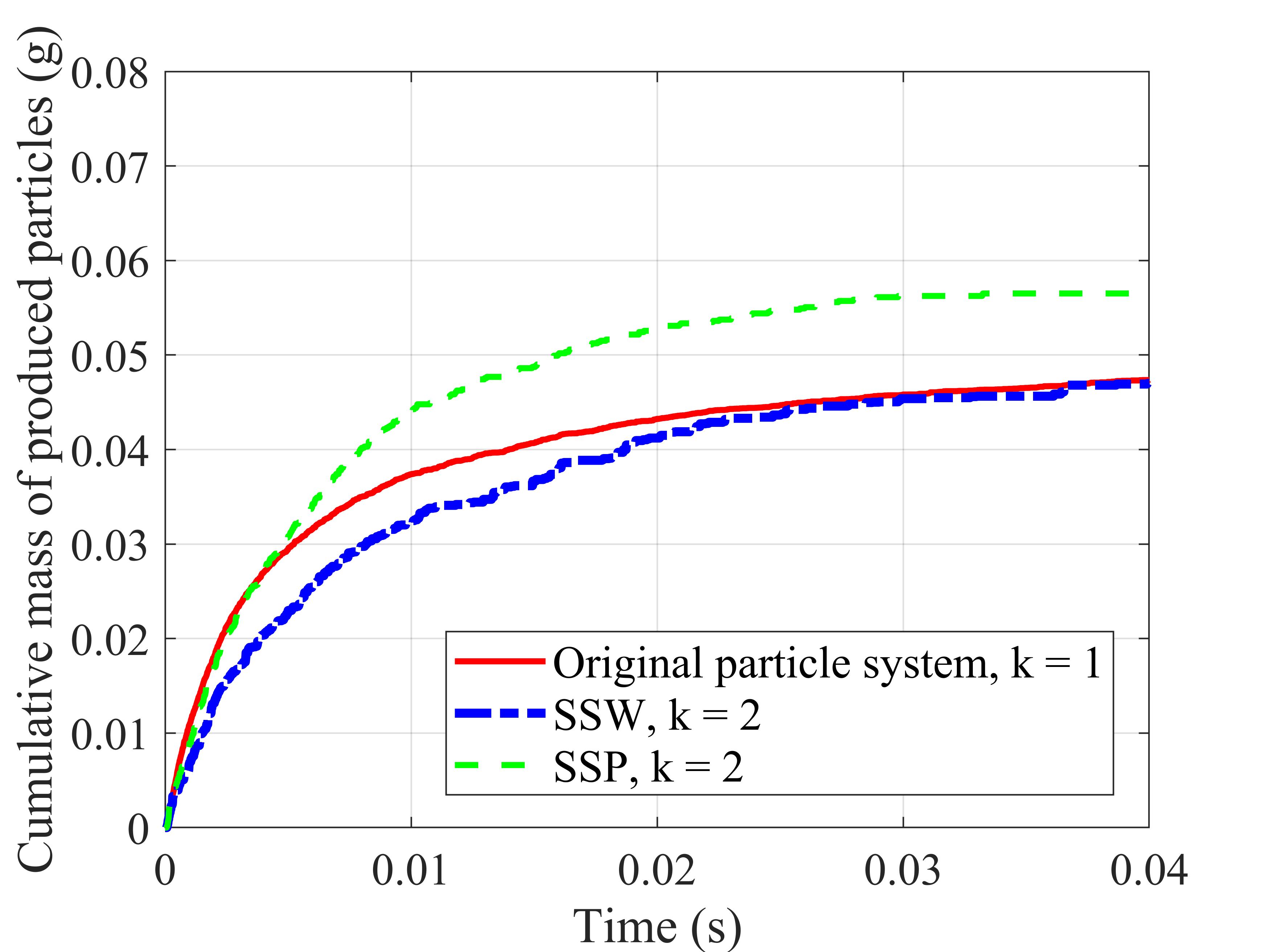}
    \caption{Comparison of the sand production rate of the original and coarse-grained particle models.}
    \label{fig:fig10}
\end{figure}

Figure \ref{fig:fig10} demonstrates the cumulative sand production rate of the original particle, SSW and SSP models. It can be seen from the figure, the sand production in all simulations occurred in a similar pattern. During the intensive sand production up to 0.005 s, SSP was the same as the original particle model, while SSW produced less sand.
From 0.005 s the sand production in the SSP model started increasing and produced more sand when transient behavior was maintained. On the other hand, sand production of the original particle and SSW models became similar by the end of simulation.  The root mean squared relative error (RMSRE) for the SSW model was 13.24 \%, and the SSP model was 19.67 \%.   By comparing RMSRE of both coarse-graining models we can state that the result of the SSW model was more accurate than the result of the SSP model in the cumulative sand production.

\begin{figure}[h!] 
    \centering
    \includegraphics[width=13cm, height=8.67cm]{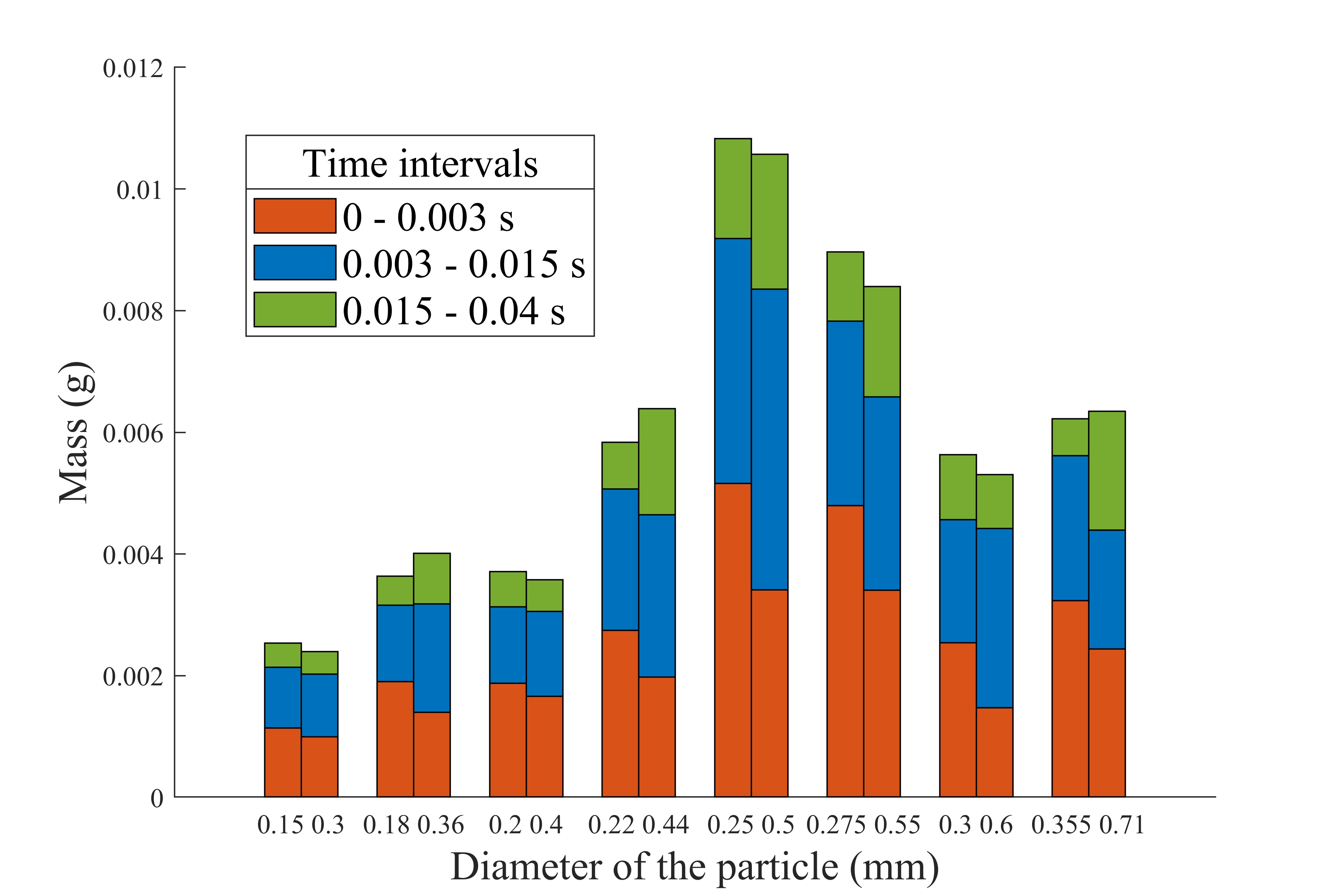}
    \caption{Comparison of the produced sand PSD of the original particle (left bar) and SSW coarse-grained (right bar) models.}
    \label{fig:fig11}
\end{figure}

Next, we compared the particle size distribution by the mass of the produced sand.  We considered only the original particle and SSW models,  in which the polydisperse behavior of the particles existed. It was mentioned before that eight different diameter groups of particles were generated in the numerical sample. Figure \ref{fig:fig11} shows the relationship between the diameter of the particles and their produced mass at different time intervals. In general, the mass of the produced particles corresponds to the initial mass ratio of the numerical sample, as shown in Table \ref{tab:table2}. At the beginning, which is $0 - 0.003$ s, the groups with  larger particles significantly differ in production mass. On the other hand, the groups with smaller particles are less differentiated compared to large particle groups. In all groups, the original system produced more particles than the coarse-grained system. This is also illustrated in Figure \ref{fig:fig10}. In the middle of the simulation, which is $0.003 - 0.015$ s, all groups showed relatively similar production results. In the final time interval, which is $0.015 - 0.04$ s, the mass of coarse-grained particles in most groups was greater than that of the original particles. Therefore, during the steady-state condition, the total mass of produced particles became equalized in both systems. 


\subsection{Comparison of the CPU time between the original and coarse-grained particle models}

\begin{figure}[h!] 
    \centering
    \includegraphics[width=10cm, height=7.1cm]{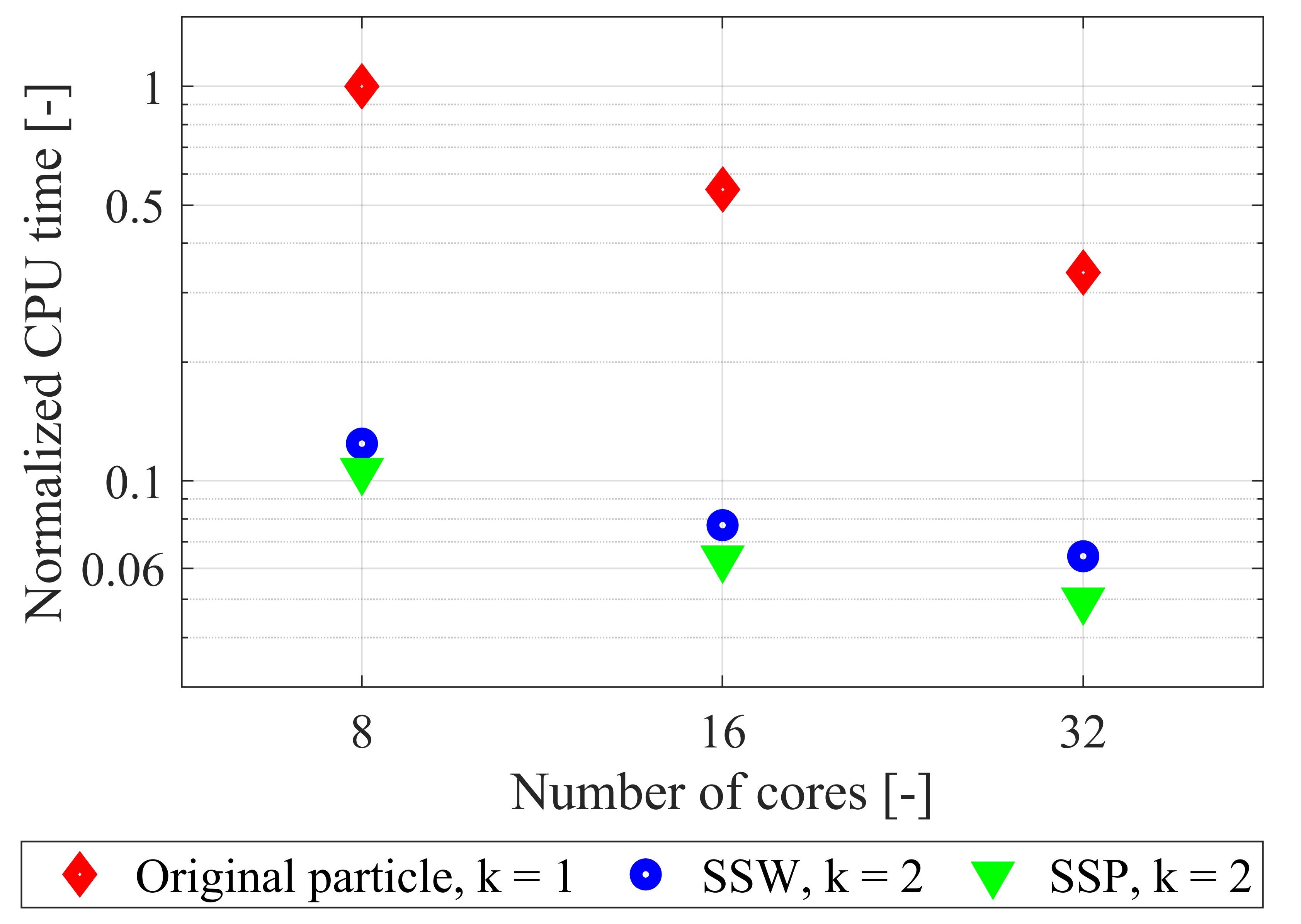}
    \caption{Comparison of the simulation CPU time results at different core numbers.}
    \label{fig:fig12}
\end{figure}

The presented results above show that the coarse-grain models can represent the original particle system in the CFD-DEM coupling model. Therefore, the simulation time of the coarse-grained models was expected to decrease significantly. In order to fairly assess the speedup of coarse-graining, the simulations were executed using parallelization in computation with 8, 16 and 32 cores. The simulation time can be defined as the CPU time of the performed simulations. We measured the CPU time of the coarse-grained simulations and compared it with the original particle system. Figure \ref{fig:fig12} demonstrates the relationship between the number of cores and normalized CPU time. In this case, the CPU time was normalized by maximum simulation time corresponding to the original particle model at 8 cores. The normalized CPU time of the simulation decreases almost gradually as the number of cores increases. We also observed the acceleration of the original system due to the increase in the number of cores. However, significant acceleration has been achieved using coarse-grained methods. Simulations performed by the SSP method showed faster acceleration at any number of cores compared to other models. The most outstanding result was shown at 32 cores, where the shortest simulation time was achieved. Such a fast speedup of the SSP method is achieved due to the monodisperse behavior of particles in the model.

Additionally, in order to understand the effectiveness of the coarse-graining methods the speedup was computed. In our case, the speedup is defined as follows:
\begin{equation}
Speedup = \frac{t_{p}}{t_{g}}   \label{eq:43}
\end{equation}
where $t_{p}$ - the CPU time of the original particle simulation and $t_{g}$ - the CPU time of the coarse-grained simulation. 

The speedup term shows how well the coarse-graining method accelerates the original particle model when the simulations are executed with the same number of cores. The calculated speedup of the simulations are provided in Table \ref{tab:table5}. As it was expected the speedup of the SSP method was higher than SSW at all core numbers due to its monodisperse behavior of the particle system. The most excellent speedup was observed on 8 cores. At this core number the SSP method has accelerated the simulation by 9.4 times. The parallelization implementation of the CFD-DEM model including the communication cost can still be improved since an increase in the number of cores led to a decrease in the speedup of coarse-grained models. 
For instance, although the simulations using coarse-graining methods were performed faster with 32 cores, they were less efficient in terms of speedup than with 8 and 16 cores. 
The CPU time of the coarse-grained simulations can be reduced by using a large number of cores.


\begin{table}[h!]
 \caption{Speedup due to the coarse-graining method}
  \centering
\begin{tabular}{ l l l l } 
\hline
\multirow{2}{4em}{Number of cores} & \multicolumn{3}{c}{Speedup} \\ 
\cline{2-4}
& Original particle  & SSW (k = 2) & SSP (k = 2) \\
\hline
 8   & 1  & 8.1 &   9.4\\
 16   & 1  & 7.1  & 8.6 \\
 32   & 1   & 5.2  &  6.8 \\
 \hline
 \end{tabular}
   \label{tab:table5}
\end{table}


\section{Conclusions}
The main aim of this work is to present the coarse-graining methods of 3D CFD-DEM model  based on the quadratic force scaling for the sanding phenomena in unconsolidated reservoirs. The derivation of the transformation from original (fine scale) to the coarse grained model is shown mathematically. The simulations were performed in the modified cohesive contact model, in which the bond breakage between particles occurred at the maximum value of the normal contact force. The original model is initially verified for the laboratory experimental data associated  with the Kazakhstan reservoir field's the particle size distribution (PSD). We have compared the results of the proposed coarse-grained models with the original model including the fluid velocity, the cumulative sand production rate, and PSD of produced sand particles. The results of coarse grain models, which has 1/8 of the particle number of the original system, are in good match with the results of the sand production rate for the original particle model. The SSW model demonstrates more accurate results compared to the SSP model. Not only the heterogeneity of the porous medium is represented in the SSW model but also PSD of the produced particles are similar to the original model result. To improve computational cost of the coarse-grained model, we have conducted numerical experiments in the parallelized setting with maximum 32 cores. 
As a result, the CPU time of the coarse-grained simulations decreased as number of cores increased. In addition, the best speedup was shown for the model with 8 cores where the CPU time of SSP model is 9.4 times faster than the CPU time of the original model. The findings also suggest that the SSW approach could be efficient for the prediction of the sand production rate accurately using polydisperse particles in the CFD-DEM system with a modified cohesive contact model.

\bibliographystyle{elsarticle-harv} 
\bibliography{references}

\end{document}